\documentclass[journal]{IEEEtran}

\usepackage{times}
\usepackage{soul}
\usepackage{url}
\usepackage[hidelinks]{hyperref}
\usepackage[utf8]{inputenc}
\usepackage[small]{caption}
\usepackage{graphicx} 
\graphicspath{ {./images/} }

\usepackage{amsmath}
\usepackage{booktabs}
\usepackage{algorithm}
\usepackage{algorithmic}
\usepackage{subfigure}  
\usepackage{color}
\def\HU#1{{\color{black}{#1}}}

\ifCLASSINFOpdf
\else
\fi

\makeatletter
\newcommand{\printfnsymbol}[1]{%
  \textsuperscript{\@fnsymbol{#1}}%
}
\makeatother

\begin{document}

\title{Multi-intersection Traffic Optimisation: 
A Benchmark Dataset and a Strong Baseline}

\author{Hu Wang,
        Hao Chen,
        Qi Wu,
        Congbo Ma,
        Yidong Li\printfnsymbol{1},
        Chunhua Shen
        \thanks{
        HW, HC, QW, CM, CS are with The University of Adelaide, Australia;
        YL is with Beijing Jiaotong University, China.
        }
        \thanks{* Corresponding author.}
        }

\markboth{02/2019
}%
{Wang et al.\ }

\maketitle

\begin{abstract}

The control of traffic signals is fundamental and critical to alleviate traffic congestion in urban areas. However, it is challenging since traffic dynamics are complicated in real-world scenarios. Because of the high complexity of the optimisation problem for modelling the traffic, experimental settings of existing works are often inconsistent. Moreover, it is not trivial to control multiple intersections properly in real complex traffic scenarios due to its vast state and action space. Failing to take intersection topology relations into account also results in inferior solutions. To address these issues, in this work we carefully design our settings and propose a new dataset including both synthetic and real traffic data in more complex scenarios.  Additionally, we propose a novel baseline model with strong performance. It is  based on deep reinforcement learning with an encoder-decoder structure: an edge-weighted graph convolutional encoder to excavate multi-intersection relations; and an unified structure decoder to jointly model multiple junctions in a comprehensive manner, which significantly reduces the number of the model parameters. By doing so, the proposed model is able to effectively deal with the multi-intersection traffic optimisation problem. Models are trained/tested on both synthetic and real maps and traffic data with the Simulation of Urban Mobility (SUMO) simulator. Experimental results show that the proposed model surpasses multiple competitive methods.

\end{abstract}


\IEEEpeerreviewmaketitle

\section{Introduction}

Nowadays, traffic congestion has become a practical but challenging issue in our daily lives. It not only costs a significant amount of driver's waiting time, but also causes air pollution. To address these issues, expanding road networks is a solution, but it is not always feasible due to various constraints. Developing new intelligent algorithms for traffic control offers an alternative but more economical choice.

Traffic optimisation is an active research area, where many existing works have been proposed for efficient traffic signal control \cite{li2014survey, DBLP:journals/corr/BalujaCS17, covell2015micro, liang2018deep}. These works can be roughly classified into two categories: predefined rule-based adaptive methods; learning-based methods such as hill-climbing \cite{DBLP:journals/corr/BalujaCS17} or deep reinforcement learning (DRL) based methods which attempt to fit the dynamic fluctuation of traffic flows \cite{liang2018deep}. However, in the aforementioned works, the simulation settings and datasets are usually inconsistent. Additionally, many current works consider traffic dynamics of single or few junctions. These methods show promising results under the simplified situations, but it may not generalise well to real traffic dynamics.

In order to tackle these issues, here we carefully design our settings/data and propose a strong baseline---the Edge-weighted Graph convolutional and unified structural Reinforcement Learning (EGU-RL) model. Our proposed EGU-RL model is able to control large-scale traffic junctions with a unified central agent. It can further take advantage of relations between junctions with edge-weighted graph convolutional networks by taking the spatial and distance information into account. The main contributions of this work are as follows.
\begin{itemize}
\item We carefully design our settings and propose a new dataset containing one synthetic map and multiple real maps from the city of Manhattan and Suzhou. Along with these maps, both synthetic and real traffic flows in comprehensive and complex scenarios are provided.

\item We introduce a grouped multi-agent reinforcement learning  baseline model  to group neighbouring intersections, which is able to alleviate the drawbacks of exponentially increased action dimensions.

\item 
We propose 
a novel and strong baseline model. We introduce a reinforcement-learning model with an edge-weighted graph convolutional encoder and a unified structure decoder to optimise multi-intersection flows, as well as to avoid the challenges caused by the large state/action dimensions. Consequently, the traffic signal control agent is more likely to gain a good understanding of the overall environment dynamics. Moreover, the proposed EGU-RL model contains fewer parameters and is end-to-end trainable. Our experimental results show the effectiveness of our model working in complex multi-interaction scenarios. Performance achieved by the proposed method surpasses multiple comparing methods in most cases.
\end{itemize}

\section{Related Work}

Recently, traffic optimisation has been extensively discussed. Traditional methods use fixed rules and detector-based methods to control the traffic lights. Detector-based methods (also referred to as actuated methods) leverage detectors placed underground with predefined rules to control traffic signals. Thus, they have more adaptability when comparing to fix-rule methods. Recently, many learning-based and data-driven approaches, i.e., evolutionary algorithms, have been adopted to effectively adapt traffic dynamics. The combination of learning-based methods and predefined control frameworks are also proposed by researchers, such as the auction-based methods \cite{covell2015micro,schepperle2008auction}. The work in \cite{DBLP:journals/corr/BalujaCS17} incorporates an auction-based model with Next-Ascent Stochastic Hill-climbing (NASH) optimisation to control traffic lights.

The control of traffic lights can be modelled as a Markov Decision Process (MDP) \cite{ritcher2007traffic}. As such, we can exploit reinforcement learning. A reinforcement learning agent takes the observation of current traffic conditions as its states. Following certain policies, the agent decides the next round traffic signals to switch as its actions. Then, corresponding rewards of taking an action will be returned, after which the agent updates its model accordingly. The agent will always try to maximise the accumulated expected reward as much as possible. The aforementioned process forms a closed loop that enables the agent to learn the ``correct'' action distribution under a particular circumstance. With great progress achieved by deep learning techniques, many reinforcement learning algorithms combined with deep neural networks have been proposed, i.e., Deep-Q-Network \cite{mnih2015human}, Actor-Critic \cite{mnih2016asynchronous} and Proximal Policy optimisation (PPO) \cite{schulman2017proximal, heess2017emergence}, leading to significantly improved results. They can generally be categorised into value-based methods and policy-based methods. The application of deep reinforcement learning on traffic optimisation has also been adopted by \cite{liang2018deep, genders2016using, richter2007natural, richter2006learning, mannion2016experimental}. Tong et al.\  \cite{pham2013learning} presented a preliminary comparison between reinforcement learning approaches and distributed constraint optimisation approaches for traffic signal control.

However, for multi-intersection control, because of the high complexity of modelling this problem, experimental settings are usually different among existing works. Most works simplify the situation to single or few junctions, which is largely different from real situations, rendering these approaches less useful for real-world applications. 

Moreover, these methods need to encode vast state and exponentially increased action spaces, leading to an inefficient learning procedure. Some works incorporate multi-agents methods \cite{arel2010reinforcement,liu2018deep,wang2020large,chu2019multi,wu2020multi} to alleviate this issue.
In practice, tuning hyper-parameters of a DRL model  is not a trivial work and it is even harder to tune that of a multi-agent model. If without effective communication between agents, a multi-agent system may suffer from unstable training easily and falls into sub-optimal solutions. This has been theoretically proven by research in game theory, among which the prisoner's dilemma is a typical example. Authors of \cite{nishi2018traffic} combined graph convolutional networks with reinforcement learning. The policies of each intersection are learned simultaneously but independently with an individual network per junction. It is worth noting that this work simplifies the environment to six intersections. It remains unclear how sensitive of the results presented in \cite{nishi2018traffic} with respect to the large number of parameters, as multiple isolated agents do not have communications to each other. It also increases the risk of overfitting due to its large number of parameters. In \cite{wei2019colight}, a graph attention network was proposed for the traffic optimisation task. Nevertheless, multiple intersections are processed individually and separately which increases the difficulties of model optimisation process due to the model is updated for every individual traffic signal.

Here, we propose a novel and strong baseline model to exploit relations among junctions by encoding their spatial and distance information with edge-weighted graph convolutional networks. Additionally, we introduce an unified structure decoder to jointly model multiple intersections and avoid the issue of exponentially increased action dimension explosion. The proposed method can not only outperform existing methods, but also significantly reduce the number of parameters.

On the settings aspect, Auction-based methods \cite{covell2015micro}, the work in \cite{schepperle2008auction} and \cite{DBLP:journals/corr/BalujaCS17} attempted to minimise average travel time or total travel time of vehicles. 
The work in \cite{nishi2018traffic} adopts a position matrix to map the locations and the speeds of vehicles within one intersection as the state, the duration change of different phases as its action space and waiting time of vehicles as the model's reward. In \cite{guo2019reinforcement}, individual vehicle’s position and speed have also been used as the state. However, the reward is defined to minimise the total queue length of incoming lanes. The work in \cite{mousavi2017traffic} uses the raw images or snapshots of a graphical simulator as the input state. Besides the different state and reward design, experiments only on synthetic data were conducted by most of the existing models. Thus, the settings and data provided by existing works are vastly inconsistent. Towards this end, we carefully design our settings and propose a new dataset which contains one synthetic map and several real maps from the city of Manhattan and Suzhou along with real traffic data.

\section{Settings and Dataset}

\begin{figure}[t!]
\centering
\includegraphics[width=0.35\textwidth]{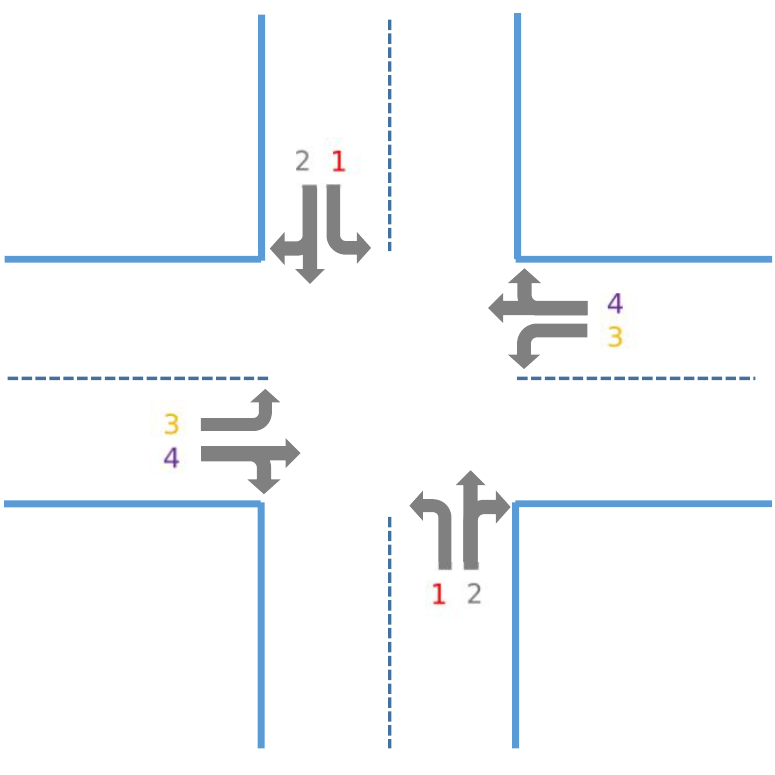}
\caption{An example of traffic signals (right-side driving). As shown in the figure, the numbers represent different phases. In each round, phases are switched from ``1'' to ``4'' sequentially to allow different traffic streams to pass the junction without collision.}
\label{fig:four_phases}
\end{figure}

Traffic light signals consist of multiple phases. The number of traffic light phases is an indicator of how many traffic streams are allowed to occupy the road in one round. For instance, in practice four-phase traffic signals are commonly seen traffic signals for crossroads. The intuitive illustration of a four-phase traffic signal is shown as Fig. \ref{fig:four_phases} (taking right-side driving for example). Phase ``1'' and ``3'' represent left-turn of N-S and E-W directions respectively. Phase ``2'' and ``4'' stand for straight/right-turn of N-S and E-W directions respectively. In each round, phases are turned from ``1'' to ``4'' sequentially to allow different traffic streams to pass the junction without collision.

For one traffic light, each phase controls one or several non-conflicting lanes. In each phase, ``green'', ``yellow'' and ``red'' signals are flashed orderly for controlled lanes. While the phase of a traffic light is changing, some ``red'' lanes will be switched to ``green''. At the same time, some other lanes are switched from ``green'' to ``red'', during which ``yellow'' signals are inserted as intermediate phases for warning purpose. The ``yellow'' signals delay the effect of phase switching that complicates the problem even more.

\subsection{Formulation as a Markov Decision Process}

The aforementioned traffic control task can be modelled as a Markov Decision Process (MDP). A typical Markov Decision Process can be described as a four-tuple $\left<S, A, R, P \right>$, where $S$ stands for the set of partially observed states. $A$ is the set of all possible actions. $R$ represents the collection of the rewards received by taking certain actions. $P$ means the transition probability set among different states. The traffic optimisation problem formulation is described in detail subsequently.

\subsubsection{State Space}

To realistically simulate the environment, vehicles' starting/stopping acceleration and drivers' reaction time are taken into consideration. The goal of state spaces design is to encode observations of agents as comprehensive, practical and simplified as possible. Some existing works treat raw images as states \cite{mousavi2017traffic}. Instead of raw images, different states have been designed in our settings, making model training more efficient.

\def\Thresh{{\rm Thresh}}
\def\VEH{{\rm VEH}}

The number of waiting vehicle state $S_{num}$ is the concatenation of each single lane waiting vehicle number, which is shown as Eq. \eqref{eqn:state_car_num_total} and Eq. \eqref{eqn:state_car_num}. It counts the number of vehicles if their velocity is below a certain threshold.
\begin{equation}\label{eqn:state_car_num_total}
    S_{num} = \{S_{num}^1, \, S_{num}^2,\, ..., \, S_{num}^m\}
\end{equation}
\begin{equation}\label{eqn:state_car_num}
    S_{num}^j = \sum_i veh_i^t, \forall veh_i^t \in \VEH_t, \mbox{ if } v(veh_i^t)
    \leq \Thresh,
\end{equation}
where $m$ is the total lane number.
$S_{num}^j$ represents number of waiting vehicle for lane $j$.
$\Thresh$ is a vehicle speed threshold.
$\VEH_t$ is the set of vehicles within controlled lanes at time $t$.
$veh_i$ represents individual vehicle.
$v(x)$ is the function to retrieve the speed of a vehicle.

However, $S_{num}$ fails to take the overall waiting time of vehicles into account. Thus, we calculate the total vehicle's accumulative waiting time $S_{wt}$ as well from an orthogonal angle.
Similar to the number of waiting vehicle state, the total accumulative waiting time state $S_{wt}$ is the concatenation of accumulative waiting time of individual lanes:
\begin{equation}\label{eqn:state_wt_total}
    S_{wt} = \{S_{wt}^1, S_{wt}^2, ..., S_{wt}^m\}
\end{equation}
\begin{equation}\label{eqn:state_wt}
    S_{wt}^j = \sum_i {\rm WT}(veh_i^t), \forall veh_i^t \in  \VEH_t
\end{equation}
where $ {\rm WT}(veh_i^t)$ is the function to retrieve vehicles' accumulative waiting time. Thus, the state at time step $t$ is the concatenation of $S_{num}$ and $S_{wt}$:
\begin{equation}
    S_t = \{S_{num}, S_{wt}\}
\end{equation}

\begin{figure}[t]
\centering
\includegraphics[width=0.4\textwidth]{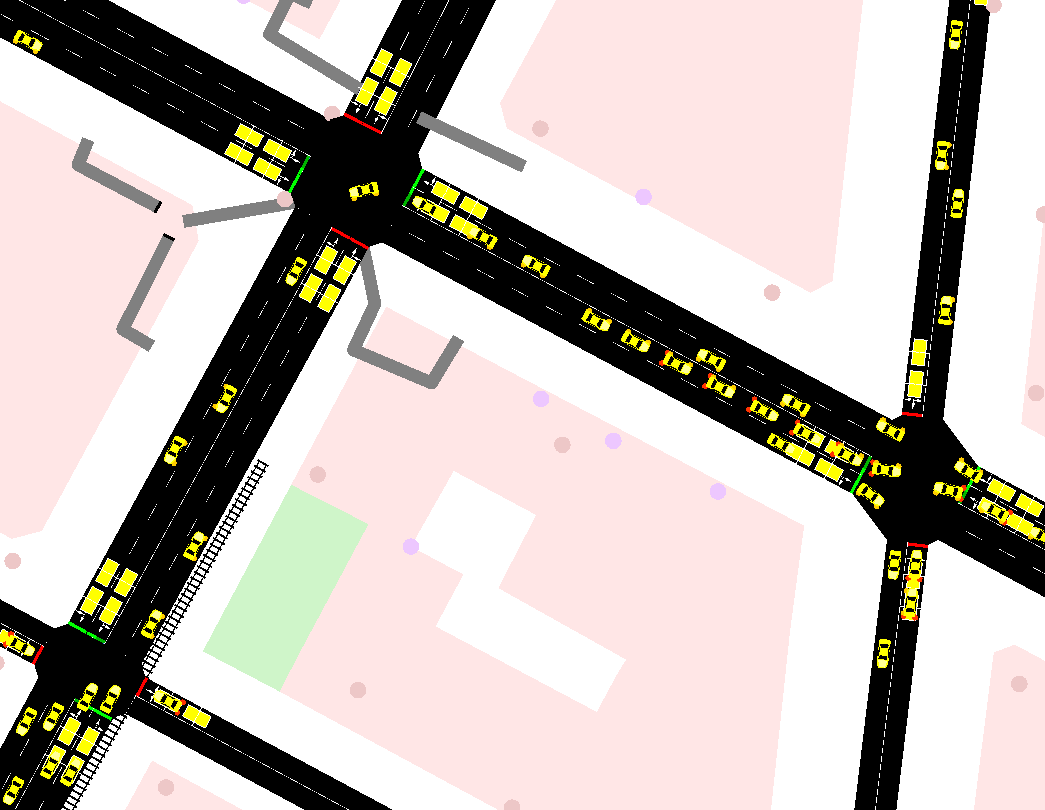}
\caption{\HU{An example of traffic optimisation scenario. Based on the traffic scenarios, we design two states: (1) total number of waiting cars and (2) total vehicle's accumulative waiting time.}}
\label{fig:states}
\end{figure}

In this case, both the number and accumulated time of waiting vehicles are now taken into account. It empowers the model to treat vehicles with different waiting time differently. A typical example of traffic optimisation scenario is shown as Fig.~\ref{fig:states}.

\subsubsection{Action Space}

As described in the previous section, the number of traffic light phases is an important indicator to show how many traffic streams are allowed to occupy the road. Instead of designing a hard constraint, i.e., phase switch sequence and duration, we set each phases to be selected as actions. In this case, the controller is able to switch between phases freely to maximise the learning ability of models. More concretely, for a four-phase traffic signal controller, its actions space contains four elements which are corresponded to four different phases. If the initial phase is NS-Right-Turn (phase 1 in Fig.~\ref{fig:four_phases}) and the controller tends to switch to Left Turning (phase 2 in Fig. \ref{fig:four_phases}), an action $1\rightarrow2$ is required.

The one-hot vector is the most commonly adopted encoding method. However, if we would like to control multiple traffic signals with a central agent, one-hot action vector representation is no longer an ideal solution. Because the phase combinations of different traffic lights are required to be encoded into one unified action space to ensure there is no conflicts and ambiguity, which incurs an exponentially increased action space.

In the design of state and action space, we propose an ``group agent" strategy. In this strategy, intersections close to each other will be grouped together and they are controlled by a shared agent. We call it grouped multi-agent reinforcement learning model. With this strategy, actions can still be encoded as one-hot vectors since the number of traffic lights within a group is not significantly huge. For each action $a$ satisfying $\forall a \epsilon A$. In term of the agent action space design, we have also proposed a unified structure decoder to output actions of all traffic lights at the same time, which has different action space definition and will be elaborated in detail in section IV.

\def\WT{{\rm WT}}

\subsubsection{Reward Design}

Reward is the feedback from the environment to indicate the effectiveness of performed actions. In our settings, we define the total accumulative waiting time improvement as an instant reward at time step $t$, which is denoted as:

\begin{align}
    R_{wt} = \sum_i \WT(veh_i^{t-1}) - \sum_j \WT(veh_j^t)
\end{align}

Intuitively, if a ``correct'' action is performed, the total accumulative waiting time from the last step will be reduced significantly, so a large $R_{wt}$ will be received; otherwise, a small $R_{wt}$ will be returned.

\subsubsection{Evaluation Metric}

In the proposed traffic optimisation setting, the total accumulative waiting time cost is adopted as our evaluation metric to measure different algorithms. We define the total accumulative waiting time cost at each step $t$ as:
\begin{equation}\label{eqn:cost_wt}
    Cost_{wt} = \sum_i \WT(veh_i^t), \forall veh_i^t \in \VEH_t
\end{equation}
As stated in the previous section, $ \WT(veh_i^t)$ also represents the accumulative waiting time of a certain car within control.

Eq.~\eqref{eqn:cost_wt} reflects our life experience: if an appropriate action is perform, the accumulative waiting time cost at time step $t$ is going to drop. Accordingly, a small $Cost_{wt}$ will be returned. Otherwise, the model will receive a larger accumulative waiting time cost. Therefore, the goal of our task is to minimise the total accumulative waiting time cost. If the cost is minimised, it means that a promising model is learned.

\subsection{Data}

In this subsection, we present the details of the data, including two parts: (1) traffic maps and traffic flow data. (2) statistics of these data.

\subsubsection{Maps and Traffic Data}

`Env' is a synthetic map with fifteen intersections. `Man1' and `Man2' are real traffic maps of Manhattan ares New York and each of them contains twenty-two intersections. `Suzhou1' is a ten-intersection real map with real traffic flows collected from Suzhou China. `Suzhou2' is a real scene from Suzhou as well with twelve intersections.

In our simulation, the aforementioned waiting car number state is retrieved from road sensors. Additionally, two detect loops are placed at the first five and ten meters respectively of each lane to retrieve the accumulative waiting time state. All of our experiments are trained and evaluated in five different scenes. In the map Suzhou1, the real traffic flow was generated by tracking the from points and to destinations of real vehicles. Besides Suzhou1, three different traffic flows with different busy conditions are generated to keep data diversity. The details of our maps and data are shown in Table \ref{tab:maps_data}.

\begin{table}[t]
\centering
\caption{Maps and traffic data. `Env' is a synthetic scene with fifteen intersections. `Man1' and `Man2' are scenes of Manhattan blocks of New York and each of them contains twenty-two intersections. `Suzhou1'  is a ten-intersection real map with real-world traffic flows collected from Suzhou, China. `Suzhou2' is a scene from the blocks of Suzhou with twelve intersections. \HU{'Synth' represents the synthetic traffic data.}}
\scalebox{1.15}{
\begin{tabular}{@{}l lllll@{}}
\toprule
Scenes                & Env      & Man1 & Man2 & Suzhou1 & Suzhou2   \\ \midrule
Maps             & Synth & Real       & Real       & Real    & Real      \\
Traffic & Synth & Synth  & Synth  & Real    & Synth \\
Inters Num  & 15        & 22         & 22         & 10      & 12        \\ \bottomrule
\end{tabular}}
\label{tab:maps_data}
\end{table}

\subsubsection{Simulation and Statistics of Data}

\begin{figure}
\centering
\includegraphics[width=0.5\textwidth]{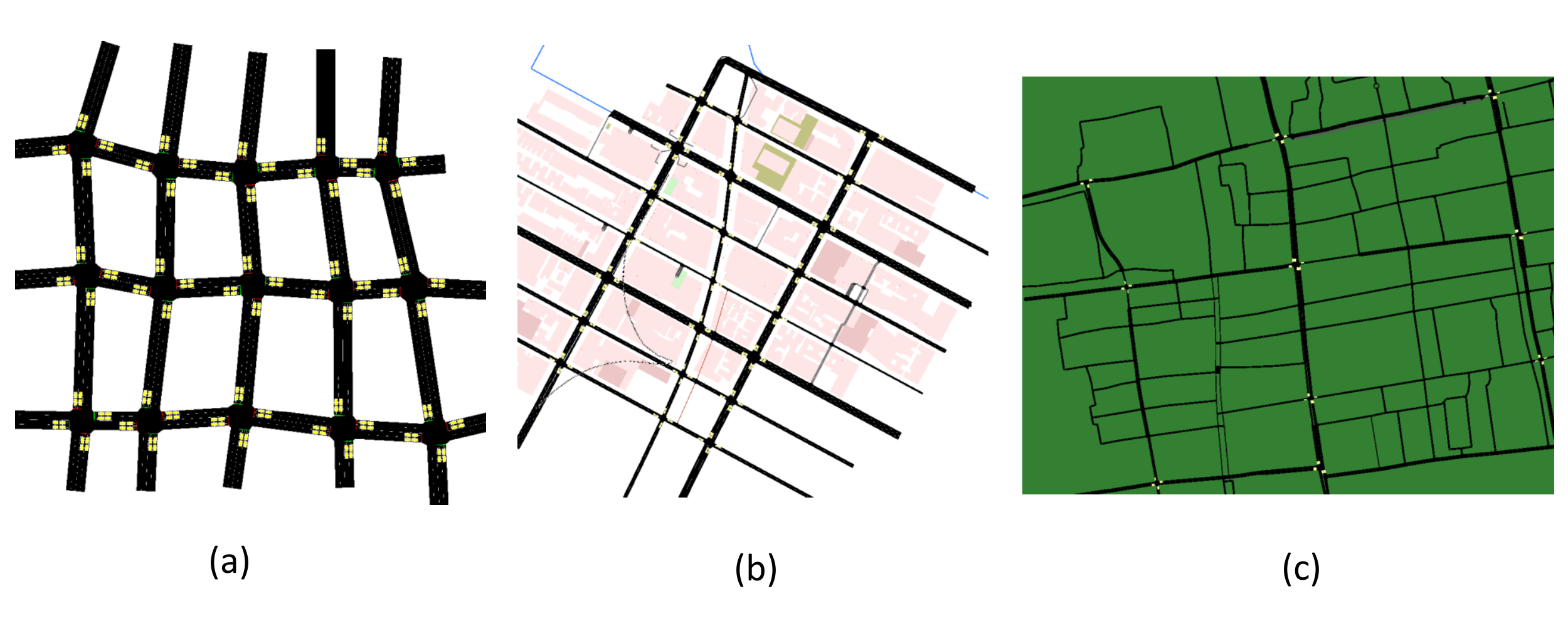}
\caption{Some maps of the proposed dataset. In the figure, the (a) is the synthetic Env map; (b) is the Manhattan2 map and (c) is the Suzhou1 map.}
\label{fig:traffic-maps}
\end{figure}

Compared with existing works, more intersections have been included in our data, ranging from ten to twenty-two intersections. Some maps of the proposed dataset can be found in Fig. \ref{fig:traffic-maps}. Simulation of Urban MObility (SUMO) \cite{SUMO2018} has been used as our simulator throughout our experiments. In order to simulate closely to real situations, vehicles' starting/stopping acceleration and drivers' reaction time have been considered as well. Settings can be found in Table \ref{tab:sim_settings}.

We split our traffic scenario into 1000 simulation steps for training. For the testing phase, we evaluate different models on seen simulation steps and further extend it to unseen ones for generalisation evaluation.

\HU{W.r.t. the generation of traffic flows. The real traffic flow was generated by tracking the from points and to destinations of real vehicles. The other synthetic traffic flows were generation with SUMO. We generated vehicles with equidistant departure times and period equals to 0.8, 1.0 and 1.2 to form the three flows of a scene. The weight of fringe edges is set to 10. The number of attempts for finding a trip which meets the distance constraints is set to 100.}

\begin{table}[t]
\centering
\caption{Our simulation setting. `Suzhou1' is a ten-intersection real map with real-world traffic flows dynamics collected from Suzhou China. For other scenes, in order to have a simulation close to real scenarios, vehicles' starting/stopping acceleration and drivers' reaction time (the time delays of drivers from observing state changes to take actions) and other important factors have been modelled.}
\label{tab:sim_settings}
\scalebox{1.25}{
\begin{tabular}{lll}
\hline
Settings  &  Values \\
\hline
Starting Acceleration         & 2.0 $m/s^2$     \\
Stopping Acceleration        & 4.5 $m/s^2$      \\
Driver Reaction Time         & 0.8 $s$      \\
Vehicle Length         & 4.5 $m$     \\
Vehicle Gap             & 1.5 $m$      \\
Speed Deviation                 & 0.2      \\
\hline
\end{tabular}}
\end{table}

\section{EGU-RL Model} 

\begin{figure*}
\centering
\includegraphics[width=0.9\textwidth]{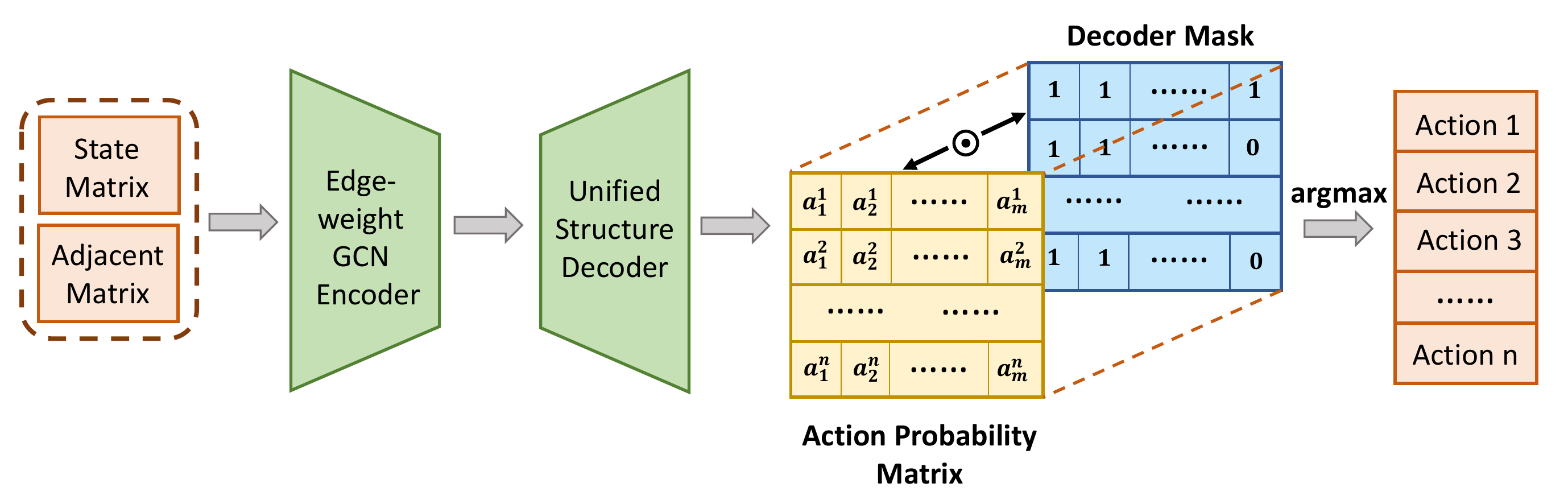}
\caption{The proposed EGU-RL Model. The figure represents our proposed EGU-RL model with the Edge-weighted Graph Convolutional Encoder and the Unified Structure Decoder.}
\label{fig:framework}
\end{figure*}

The proposed edge-weighted graph convolutional and unified structural Reinforcement Learning (EGU-RL) model contains two complementary parts: (1) An edge-weighted graph convolutional encoder to excavate positional relations among junctions. (2) A unified structure decoder to integrate the control of multiple intersections into the controller. The model structure is shown in Fig. \ref{fig:framework}.

\subsection{Edge-weighted Graph Convolutional Encoder}

Convolutional Neural Networks (CNNs) have achieved tremendous successes in most computer vision tasks. However, Standard CNN techniques are mostly designed to deal with euclidean data with grid-like structures \cite{bronstein2017geometric}. However, for the traffic optimisation task, the spatial relations of intersections are graphs that makes the problem to be non-euclidean. Graph Neural Networks \cite{kipf2016semi, scarselli2009graph, battaglia2018relational} (GNNs) offer a new point of view with a more generic framework to encode these topology node relations. Thus we further exploit cross-intersection relations using edge-weighted graph convolutional networks by taking the spatial and distance information into consideration.

For layer-wise representations, the idea of how the Graph Convolutional Networks work in the multi-intersection traffic optimisation problem is shown as below:
\begin{equation}\label{eqn:gcn}
    F^{(l+1)}=f(F^{(l)},A), l \geq 1
\end{equation}
where $F^{(l)}$ is the feature map of layer $l$, $A$ represents the adjacent matrix of all junctions. As shown above, the network structure sends the current layer's features and adjacent matrix into the network to produce features of the next layer. $F^{(0)} = X$, and $X$ is the original input of the network.

Following the previous work \cite{kipf2016semi}, we use graph convolutional nets combined with reinforcement learning to solve this optimisation problem. Additionally, we take a further step to propose an edge-weighted graph convolutional encoder to assign different weights to edges according to lane length. Because not only to know the topological relations among junctions is significant, but also to perceive the distance is crucial for the model optimisation process. It offers more information of edge importance than raw edge relations.
\begin{equation}\label{eqn:gcn2}
    f(F^{(l)},A)=\sigma (\widehat{A}F^{(l)}W^{(l)})
\end{equation}
\begin{equation}\label{eqn:gcn2_Ahat}
    \widehat{A} = normal(A+I)
\end{equation}
where $W^{(l)}$ is the weights of layer $l$ and $\sigma$ here represents the activation function in our proposed model. The adjacent matrix $A$ is defined with edge-weighted elements to attend on edges of different lengths. In this case, the elements within it is not constrained to 0 or 1. More specifically, the edge weights are the quotient of the maximum lane length divided by each lane length. Intuitively, the shorter of the length, the larger weight will be assigned. In terms of $\widehat{A}$, we follow the Graph Convolutional Networks to further normalise them row-wisely to alleviate node weights imbalance, and then add it to a identity matrix $I$, shown as Eq. \eqref{eqn:gcn2_Ahat}

\subsection{Unified Structure Decoder}

In multi-intersection traffic optimisation, if intersections are close to each other, it is desired to control them as groups instead of treating them as separated ones. It is because grouped controllers are able to perceive more information and take appropriate actions by jointly modelling traffic flows as a whole. Otherwise, if each traffic light acts isolatedly, it may lead to sub-optimal results in dealing with urban-level traffic dynamics. Thus, we propose a grouped multi-agent reinforcement learning controller to group close-related intersections for action dimension reduction. For example, if three four-phase traffic signals are grouped, action space can be presented by 64 bits with the one-hot encoding, which is exponentially increasing with the addition of controlled junctions.

Based on that, we further propose a unified structure decoder for multi-intersection control to decoder actions of multiple traffic signals in a unified manner. With this strategy, agents are able to plan and control large-scale traffic nets comprehensively since there is no internal friction to coordinate different agents. Intuitively, instead of using individual one-hot vectors for an agent as actions, the unified structure decoder is able to control all intersections with a $m$ by $n$ action probability matrix, where $n$ denotes the number of intersections and $m$ denotes the phase number of each signal (e.g., $n = 4$ for a four-phase traffic signal). In the case of the phase numbers of various traffic lights are different, the action probability matrix is multiplied with a masked matrix to ensure each signal is controlled properly.

By doing so, the vast multiple signal action size can be reduced dramatically. The state is the concatenation of all controlled intersection observations. Thus, all junctions are taken into considerations. When using one-hot action encoding, the action space of three four-phase traffic lights is 64 bits ($ 4 \times 4 \times 4$, which is exponentially increasing with the addition of controlled junctions). However, with the proposed unified structure decoder, the outputted action is a $3 \times 4$ action matrix, which is 12 bits ($4+4+4$). This is because we output them at the same time as an action matrix with the help of unified structure decoder. Thus, the model equipped with the unified structure decoder is able to deal with the dimensionality curse of exponentially increased action size. It enables EGU-RL model to control complex traffic flows with significant less parameters.

For the unified structure decoder, to let the agent have more comprehensive observations, we introduce a new state $S_{P}$ to consider current phases of all the other traffic signals.

\begin{align}
    S_{P} = \{cur\_phase_{1}, cur\_phase_{2}, ..., cur\_phase_{m}\}
\end{align}

$S_{P}$ is the concatenation of current phases of all traffic lights. $m$ denotes the number of intersections within control. So the hybrid state space becomes:

\begin{align}
    S = \{S_{N}, S_{T}, S_{P}\}
\end{align}

\subsection{Optimisation of the EGU-RL Model with Reforcement Learning}

In the aspect of reinforcement learning, rewards indicate the effectiveness of performed actions. At time $t$, the discount cumulative reward $R_{t}$ is:

\begin{align}
    \overline{R_t} = E[r_{t} + \lambda r_{t+1} + ...] = E[\sum_{i=0}^\infty\lambda^ir_{t+i}]
\end{align}

where $r_{t}$ defines reward at time $t$, $\lambda$ is the reward discount factor.

It is impractical to use the total travel time as a cost function since it cannot be calculated immediately at every time step which will dramatically slow down the model convergence. Therefore, we adopt the Temporal difference (TD) updating strategy instead of episode-wise updating strategy for the reinforcement learning learner to update the model. It achieves less training variance and is more data-efficient by combining with DQN experience replay.

Beside the total accumulative waiting time improvement reward $R_{wt}$ stated in the previous section, in order to encourage the agent not switch phases frequently, we introduce another novel unchanged reward:

\begin{align}
    R_{uc} = \sum_{i} l_{i}, \forall l_{i} \epsilon L, \mbox{ if } l_{i} \, 
    \rm is ~   unchanged
\end{align}

where $L$ represents the set of all traffic lights and $l_i$ denotes each traffic light. Intuitively, the more unchanged traffic signals, the more rewards will be returned.

Therefore the total hybrid reward function is:

\begin{align}\label{eqn:instant_reward}
    R_{t} = R_{wt} + \alpha R_{uc},
\end{align}

where $\alpha$ is a trade-off factor between two rewards. To simplify the model optimisation process, we do not heuristically set any other constraints, i.e. minimum or maximum duration of each phase.

For optimisation algorithms, we choose Double Deep Q-Networks (DQN) \cite{van2016deep} as our RL algorithms. Experience replay \cite{schaul2015prioritized} and fix-target network techniques have also been used to improve the training efficiency and stability. During training, we update our model in the temporal difference (TD) manner rather than episode-wise. With the help of experience replay, the DQN-based EGU-RL model is transferred to an off-policy RL method. Mathematically, we have:

\begin{equation}\label{eqn:dqn}
    Q(s,a,\theta') = Q(s,a,\theta) + \beta [r_t + \gamma  \max_{a'}Q(s',a',\phi ) - Q(s,a,\theta)]
\end{equation}

where $\theta'$, $\theta$ and $\phi$ are different learnable weights. $\beta$ is the learning rate. $r_t$ is the instant reward for time $t$ defined by Eq. \eqref{eqn:instant_reward} and $Q( \cdot )$ represents Q networks.

\section{Experiments}

In this section, we report experimental details and results. Besides the EGU-RL model, the competing methods include fix-rules controller, random controller, auction-based model \cite{DBLP:journals/corr/BalujaCS17}, single-agent for single-junction multi-agent reinforcement learning model (MARL-s) to control one junction with individual agent, and grouped multi-agent reinforcement learning model (MARL-g) that controls a group of traffic signals with one agent. As a competing method, the grouped multi-agent reinforcement learning model is also proposed in this paper to group spatial-closed intersections for better control. We carry out extensive experiments to study the optimal agent configuration. We empirically find that as the number of isolated agents increases, the performance drops significantly and generalises  poorly on the unseen data. In contrast, our proposed EGU-RL model is able to 
process 
large-scale junctions with a jointly modelled agent and achieves promising results. All models are trained and evaluated on various maps of Suzhou, Manhattan and synthetic scenes.

\subsection{Experimental Details}

The experiments are conducted on the platform with an Intel Xeon CPU E5-2680 and 128GB RAM. To better simulate the real scenes, many simulation factors have been considered. Those settings are shown in Table \ref{tab:sim_settings}. In the experiments, all models are implemented with PyTorch \cite{paszke2017automatic} and Simulation of Urban MObility (SUMO) simulator v0.32.0 \cite{SUMO2018}. The proposed EGU-RL model and other competing methods are trained for 500 episodes and each episode contains 1000 simulation steps. For all DQN-based methods, experience replay \cite{schaul2015prioritized} and fix target network \cite{van2016deep} techniques are used. The target network for target-Q-value retrieving has an identical structure as the inference network and the parameters of the target network will be overridden every 100 steps. For EGU-RL method, the network structure is simple. It consists of 2 hidden layers. After obtaining and concatenating observations of each junction, the state matrix is inputted to a layer with 128 neurons. Then, the outputted features will be fed to a 64-neuron layer and then output an action probability matrix. In the grouped multi-agents reinforcement learning (MARL-g) model, each agent controls three close-neighboured intersections. The learning rate of each algorithm is configured as $10^{-3}$. During training, the DQN reward discount factor $\gamma = 0.9$ is set. The agent's new action exploration rate is set to 0.5 initially, and reduce to its 0.99999 each learning step until reaches 0.01. Additionally, the buffer size for experience replay and batch size are set to be 2000 and 150, respectively. We keep all aforementioned hyper-parameters unchanged throughout 
our experiments. All of the models in our experiments are trained from scratch for a fair comparison.

\subsection{Model Performance}

The performance of different models are presented in Table \ref{tab:performance}. Each model was tested on seen 1000 simulation steps for each scene. The vehicles' total accumulative waiting time cost is used as evaluation metric (the lower the better). \HU{In the table, ``flow1'', ``flow2'', ``flow3'' are the traffic flows generated with different tempo of vehicles passing by and ``flow'' is the real traffic flow.} 

\begin{table}[t]
\centering
\caption{Cross-model accumulated waiting time consumption performance comparison in seen environments. The unit here is hours. The less accumulated waiting time consumed, the better performance of the model has. The EGU-RL method achieves the best results among all the other methods in most cases. The best results for each row are in bold.}
\scalebox{1.0}{
\setlength{\tabcolsep}{0.85mm}{
\begin{tabular}{l|cccc|cc}
\hline
Scenes          & Random & Fix  & Auction \cite{DBLP:journals/corr/BalujaCS17} & MARL-s       & MARL-g        & EGU-RL       \\ \hline
Env-flow1     & 1729   & 1230 & 661     & 283          & 362           & \textbf{222} \\
Env-flow2     & 654    & 526  & 461     & 108          & 210           & \textbf{86}  \\
Env-flow3     & 475    & 409  & 319     & 95           & 175           & \textbf{57}  \\
Man1-flow1    & 6332   & 3850 & 2249    & 3537         & \textbf{1007} & 1978         \\
Man1-flow2    & 3298   & 1753 & 1646    & 1036         & 816           & \textbf{549} \\
Man1-flow3    & 2127   & 1396 & 2099    & \textbf{313} & 1251          & 395          \\
Man2-flow1    & 4972   & 2943 & 2558    & 1197         & 1560          & \textbf{875} \\
Man2-flow2    & 2919   & 1675 & 1052    & 573          & 829           & \textbf{469}          \\
Man2-flow3    & 1530   & 1342 & 1148    & 658          & 1134          & \textbf{321} \\
Suzhou1-flow  & 883    & 657  & 409     & 302          & \textbf{151}           & 315          \\
Suzhou2-flow1 & 4461   & 2671 & 2781    & 2231         & 870           & \textbf{843} \\
Suzhou2-flow2 & 3242   & 2438 & 1959    & 794          & 514           & \textbf{368} \\
Suzhou2-flow3 & 1989   & 1090 & 1551    & 456          & 354           & \textbf{239} \\ \hline
\end{tabular}}}
\label{tab:performance}
\end{table}

From Table \ref{tab:performance}, within 13 scenes of seen environments, the EGU-RL method achieves 10 best scores. ``Auction'' represents the model proposed in the paper \cite{DBLP:journals/corr/BalujaCS17}. It combines the auction-based model with the 
Next-Ascent Stochastic Hill-climbing (NASH) algorithm to control traffic lights, which is a greedy controller. 

However, the results show the inferior performance of the greedy controller for multi-intersection control. The greedy controller does not receive ideal results probably caused by the high complexity and the fast-changing dynamics of traffic flows. The EGU-RL model can reduce the total accumulative waiting time by up to more than 80\% when compared to the fixed-scheduled controller and surpass other methods by a large margin in most cases, especially in the environments, such as Env and Suzhou2, with complex positional relationship of intersections. More specifically, EGU-RL reduces 21.6\% and 28.4\% total vehicle's accumulative waiting time compared to the second-best model in Env-flow1 and Suzhou2-flow2 respectively. \HU{In the table, our model outperforms the MARL-s model. This is due to a naive application of a multi-agent system that without effective communication and cooperation may significantly influence the model performance, even perform worse than a centralized agent. On the contrary, the proposed centralized agent does not have internal consumption and it thus can perform well under some situation, especially under the scenarios that the traffic lights are close to each other, e.g., Env-flow1 and Man2-flow1 scenarios.}

We have conducted two sets of experiments by settings and not settings the length of phases. We empirically found that vehicles will not wait infinitely if we do not set the phases length. This may be because if a vehicle waits for a long time at one intersection, the cost will be high and the agent will try to reduce the cost and release the vehicle.


We extend our model evaluation to 5000 unseen simulation steps. The total vehicles' accumulative waiting time cost is used as the evaluation metric. The generalisation evaluation is shown in Table \ref{tab:generalisation}.

\begin{table}[t]
\centering
\caption{
Accumulated waiting time consumption comparison of different models in unseen environments. Similar to seen scenarios, the unit here is hours. The less accumulated waiting time consumed, the better performance of the model has. In the generalisation test, the EGU-RL method also achieves the best results in most cases. The unit here is  hours.}
\scalebox{1.0}{
\setlength{\tabcolsep}{0.5mm}{
\begin{tabular}{l|cccc|cc}
\hline
Scenes          & Random & Fix    & Auction \cite{DBLP:journals/corr/BalujaCS17} & MARL-s       & MARL-g          & EGU-RL        \\ \hline
Env-flow1     & 25644  & 10916  & 6018    & 4620         & 32572           & \textbf{1745} \\
Env-flow2     & 4268   & 2961   & 7839    & 515          & \textbf{433}    & 456           \\
Env-flow3     & 3171   & 2494   & 2003    & 8187         & 2869            & \textbf{343}  \\
Man1-flow1    & 167626 & 149491 & 163488  & 170034       & \textbf{113034} & 180230        \\
Man1-flow2    & 115670 & 10556  & 26936   & 121190       & 9369            & \textbf{3857} \\
Man1-flow3    & 75594  & 8322   & 20403   & 6204         & 50263           & \textbf{2165} \\
Man2-flow1    & 161138 & 121285 & 95819   & 126198       & 11189           & \textbf{5791} \\
Man2-flow2    & 89933  & 9579   & 28790   & 78835        & 2897            & \textbf{2684} \\
Man2-flow3    & 35788  & 7764   & 15137   & 47274        & 6102            & \textbf{1702} \\
Suzhou1-flow  & 9024   & 3716   & 22083   & \textbf{697} & 3017            & 2995          \\
Suzhou2-flow1 & 90027  & 71567  & 67021   & 100385       & 37727           & \textbf{5788} \\
Suzhou2-flow2 & 72071  & 49926  & 82210   & 78221        & 3263            & \textbf{2393} \\
Suzhou2-flow3 & 54948  & 13270  & 20084   & 22987        & \textbf{1820}            & 1868          \\ \hline
\end{tabular}}}
\label{tab:generalisation}
\end{table}

From Table \ref{tab:generalisation}, we can clearly 
see that
the proposed EGU-RL model shows a better generalisation ability to unseen environments compared to other methods. Similar with seen experiments, within 13 scenes of seen environments, the EGU-RL method achieves 9 best scores. More specifically, EGU-RL reduces 65.1\% and 26.7\% total vehicle's accumulative waiting time compared to the second-best model in Man1-flow3 and Suzhou2-flow2 respectively. Our model can achieve better performance particularly in complex environments. This is because EGU-RL model takes junction topological relations into account and contains less parameters. Nevertheless, the naive multi-agents systems overfit easily since it is difficult for isolated agents cooperate with each other lacking in-depth communication.

\subsection{Parameter Numbers}

The EGU-RL method can not only achieve promising results, but also has fewer parameters. The
comparison of  parameter numbers is shown in Table \ref{tab:params_num}, where EGU-RL$\setminus EGE$ represents EGU-RL model without edge-weighted graph convolutional encoder.

\begin{table}[b]
\centering
\caption{The parameter number of different models. EGU-RL has the least parameters among all the other algorithms in different scenes.}
\scalebox{1.1}{
\setlength{\tabcolsep}{1mm}{
\begin{tabular}{l|ccccc}
\hline
Scenes          & Env            & Man1      & Man2      & Suzhou1         & Suzhou2         \\ \hline
MARL-s & 412.92k         & 542.38k         & 587.57k         & 274.61k         & 327.64k         \\
MARL-g         & 279.04k         & 349.09k         & 405.85k         & 162.04k         & 218.59k         \\
EGU-RL$\setminus$EGE         & 169.33k         & 185.52k         & 230.32k         & 114.26k         & 132.96k         \\
EGU-RL          & \textbf{27.78k} & \textbf{27.78k} & \textbf{27.78k} & \textbf{39.94k} & \textbf{30.34k} \\ \hline
\end{tabular}}}
\label{tab:params_num}
\end{table}

A model may have different parameter numbers in different environments due to various input and output sizes. From Table \ref{tab:params_num}, MARL-s has the most parameters since an individual agent controls each intersections, followed by grouped multi-agent reinforcement learning model and w/o EGE model. Our proposed EGU-RL method has the least parameters and saves more than 85\% of parameters compared to MARL-s method.

\subsection{Inference Time}

What is more, we also conduct experiments to measure the actual inference time of the EGU-RL model. We fetch the inference time by averaging the results of 10 times on the platform of Intel i7-8700 CPU (without GPU acceleration). The results are shown in the Table \ref{tab:infer_time}.
\begin{table}[t]
\centering
\caption{The inference time of EGU-RL model in different scenes. The unit is in ms.}
\scalebox{1.2}{
\setlength{\tabcolsep}{1mm}{
\begin{tabular}{l|ccccc}
\hline
Scenes         & Env   & Man1  & Man2  & Suzhou1 & Suzhou2 \\ \hline
Inference Time & 0.396 & 0.451 & 0.467 & 0.387   & 0.391   \\ \hline
\end{tabular}}}
\label{tab:infer_time}
\end{table}

As shown in the table, in the synthesis 15-intersection simulated environment Env, the inference time is 0.396 ms. In Man1 and Man2, due to 22 intersections are contained in each scene, the inference time is slightly slow, which are 0.451 ms and 0.467 ms respectively.
In Suzhou1 real map/flows environment, our EGU-RL model is able to predict the next action in 0.387 ms. In Suzhou2, the inference time is 0.391 ms. The results of the table show the general robustness of the model and there is no latency issue. Thus, the proposed EGU-RL model is capable of dealing with large-scale traffic data.

\subsection{Ablation Study}

To evaluate the effectiveness of individual components in our model, we further conduct ablation studies. Table \ref{tab:ablation study} shows the performance influence of the edge-weighted graph convolutional encoder and the unified structure decoder in seen simulations.

\begin{table}[b]
\centering
\caption{The performance of ablation variants in seen environments.
In the table, ``EGE'' denotes edge-weighted graph convolutional encoder and ``USD'' represents unified structure decoder. ``w/o $-$'' means the EGU-RL variant that removes the the corresponding module from EGU-RL.
}
\setlength{\tabcolsep}{0.5mm}{
\begin{tabular}{l|ccc|c}
\hline
Scenes          & w/o USD & w/o EGE & w/o EW & EGU-RL        \\ \hline
Env-flow1     & 387                   & 332                   & 275                   & \textbf{222}  \\
Env-flow2     & 131                   & 142                   & 107                   & \textbf{86}   \\
Env-flow3     & 262                   & 71                    & 105                   & \textbf{57}   \\
Man1-flow1    & 2308                  & 2053                  & \textbf{1563}                  & 1978 \\
Man1-flow2    & 4107                  & 760                   & 793                   & \textbf{549}  \\
Man1-flow3    & 534                   & 880                   & 475                   & \textbf{395}           \\
Man2-flow1    & 1461                  & 968                   & \textbf{745}                   & 875  \\
Man2-flow2    & 667                   & \textbf{381}          & 446                   & 469           \\
Man2-flow3    & 365                   & 777                   & 340                   & \textbf{321}  \\
Suzhou1-flow  & 344                   & \textbf{120}          & 451                   & 315           \\
Suzhou2-flow1 & 1485                  & 963                   & 864                   & \textbf{843}  \\
Suzhou2-flow2 & \textbf{267}          & 521                   & 509                   & 368           \\
Suzhou2-flow3 & \textbf{193}          & 303                   & 255                   & 239           \\ \hline
\end{tabular}}
\label{tab:ablation study}
\end{table}

From Table \ref{tab:ablation study} we can see that, without USD or EGE, the results experience a significant drop. ``w/o USD'' model only outperforms other two methods on Suzhou2-flow2 and Suzhou2-flow3 scenes, while beaten by its counterparts in other scenes. ``w/o EGE'' only performs well on two scenes. It is because our proposed EGU-RL model is able to not only excavate multi-intersection spatial relations with edge-weighted graph convolutional encoder, but also jointly model all controlled junctions with the unified structure decoder.

\begin{table}[htp]
\centering
\caption{The performance of ablation variants in unseen environments.}
\setlength{\tabcolsep}{0.5mm}{
\begin{tabular}{l|ccc|c}
\hline
Scenes          & w/o USD       & w/o EGE         & w/o EW & EGU-RL        \\ \hline
Env-flow1     & 5296          & 2109            & \textbf{1717}   & 1745 \\
Env-flow2     & 659           & 523             & 615    & \textbf{456}  \\
Env-flow3     & 3451          & 346             & 634    & \textbf{343}  \\
Man1-flow1    & 151606        & \textbf{123722} & 143547 & 180230        \\
Man1-flow2    & 138592        & 4337            & 4410   & \textbf{3857} \\
Man1-flow3    & 32345         & 47575           & 2675   & \textbf{2165} \\
Man2-flow1    & 133800        & 7495            & 7371   & \textbf{5791} \\
Man2-flow2    & 4617          & 3247            & 2745   & \textbf{2684} \\
Man2-flow3    & 1938          & 2226            & 2078   & \textbf{1702} \\
Suzhou1-flow  & \textbf{2225} & 5358            & 6546   & 2995          \\
Suzhou2-flow1 & 36794         & 22073           & \textbf{4903}   & 5788 \\
Suzhou2-flow2 & \textbf{2039} & 5519            & 3705   & 2393          \\
Suzhou2-flow3 & \textbf{1267} & 1789            & 3109   & 1868          \\ \hline
\end{tabular}}
\label{tab:ablation study generalisation}
\end{table}

We also carry out  experiments on 5000 unseen environments to validate the generalisation ability of different components, shown as Table \ref{tab:ablation study generalisation}. ``w/o USD'' performs poorly on most of the scenes with complex traffic flows. However, ``w/o EGE'' and EGU-RL perform relatively better, since the unified structure decoder integrates all intersections to predict actions in a more comprehensive manner. Models equipped with the unified structure decoder has remarkably fewer parameters and will not overfit easily. The edge-weighted graph convolutional encoder takes crucial cross-intersection relations into account to further boost the performance. It is easier for our network to learn complex interactions of multiple intersections because the actions are jointly modelled.


\begin{figure}[b]
\centering
\includegraphics[width=0.5\textwidth]{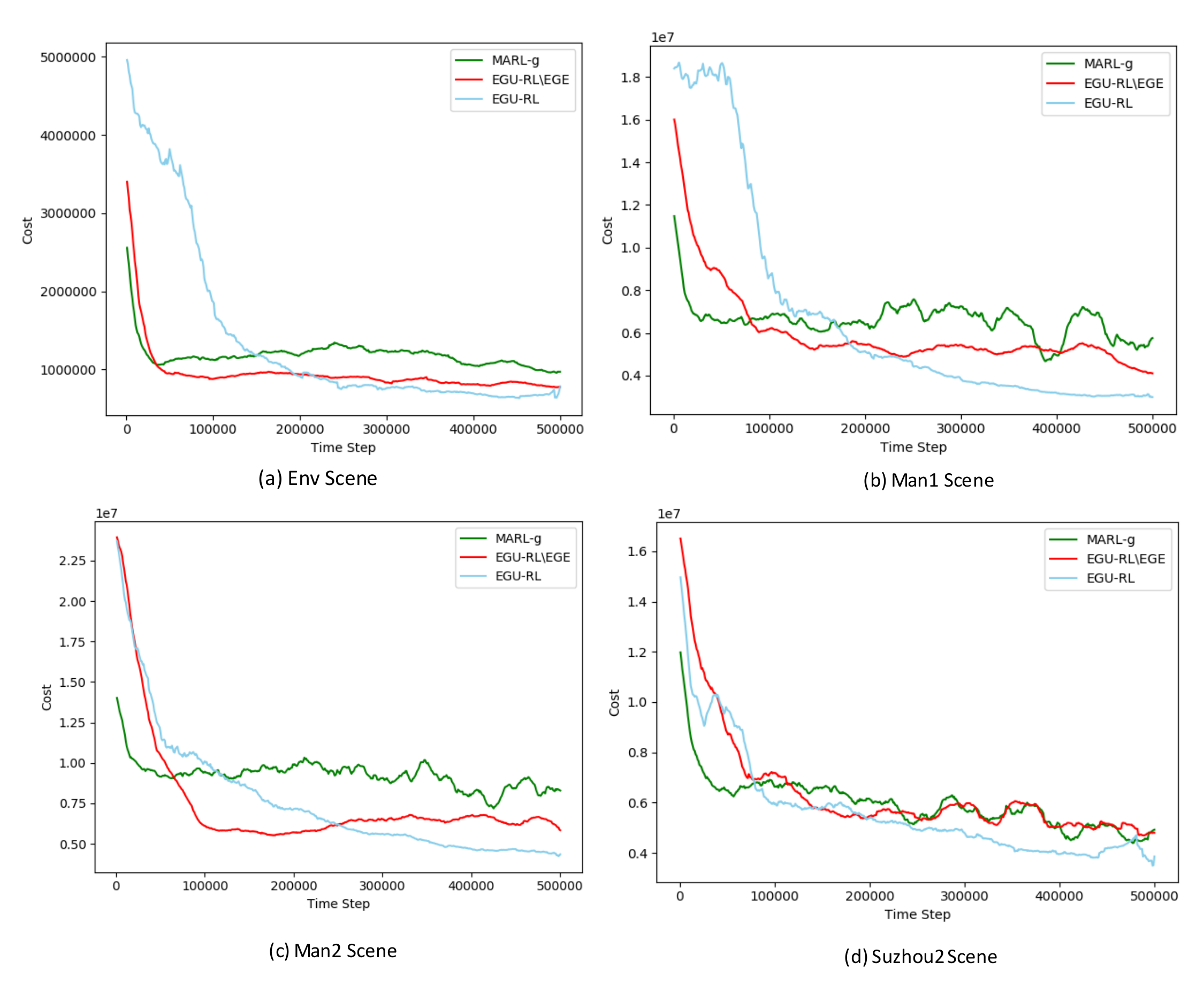}
\caption{Convergence comparison of different models on total accumulative waiting time cost (the lower the better) under different scenes. The EGU-RL algorithm surpasses MARL-g and w/o EGE algorithms and achieved the lowest accumulative waiting time costs in the end.}
\label{fig:convergence_cost}
\end{figure}

\begin{figure}[htp]
\centering
\includegraphics[width=0.5\textwidth]{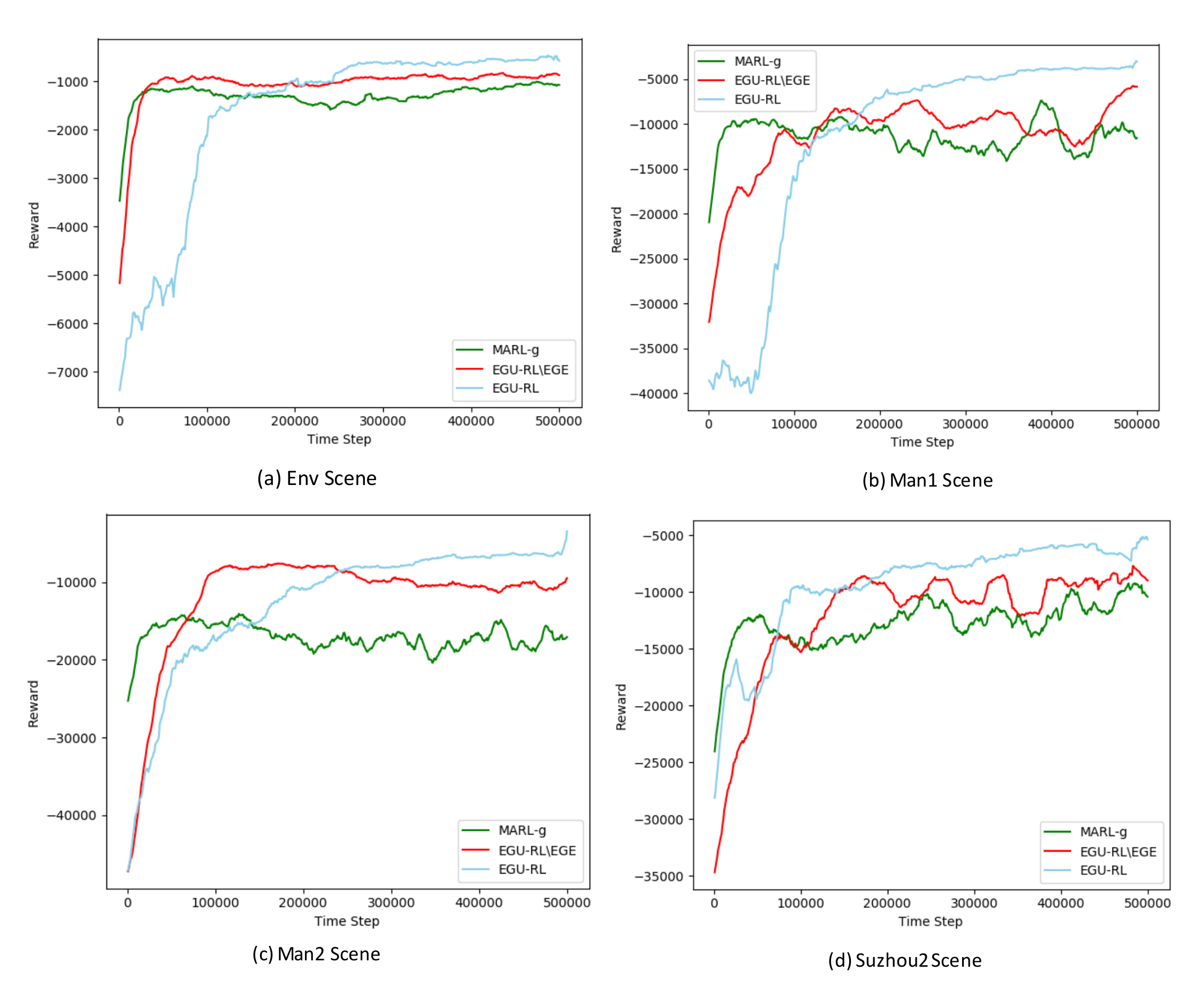}
\caption{Convergence comparison of different models on total rewards (the higher the better) under different scenes. EGU-RL algorithm surpasses MARL-g and w/o EGE algorithms and gets the highest reward in the end.}
\label{fig:convergence_r}
\end{figure}

The cost convergence comparison is shown in Figure \ref{fig:convergence_cost}. The EGU-RL algorithm surpasses the MARL-g method and w/o EGE method and achieves the lowest cost of total accumulative waiting time in the end. Moreover, our proposed EGU-RL algorithm shows much stabler training curves than the other two models. The corresponding rewards obtained by different models are shown in Fig. \ref{fig:convergence_r}. Similarly, our proposed EGU-RL model surpasses MARL-g and w/o EGE models in the end by a large margin.

\begin{figure}[htp]
\centering
\includegraphics[width=0.5\textwidth]{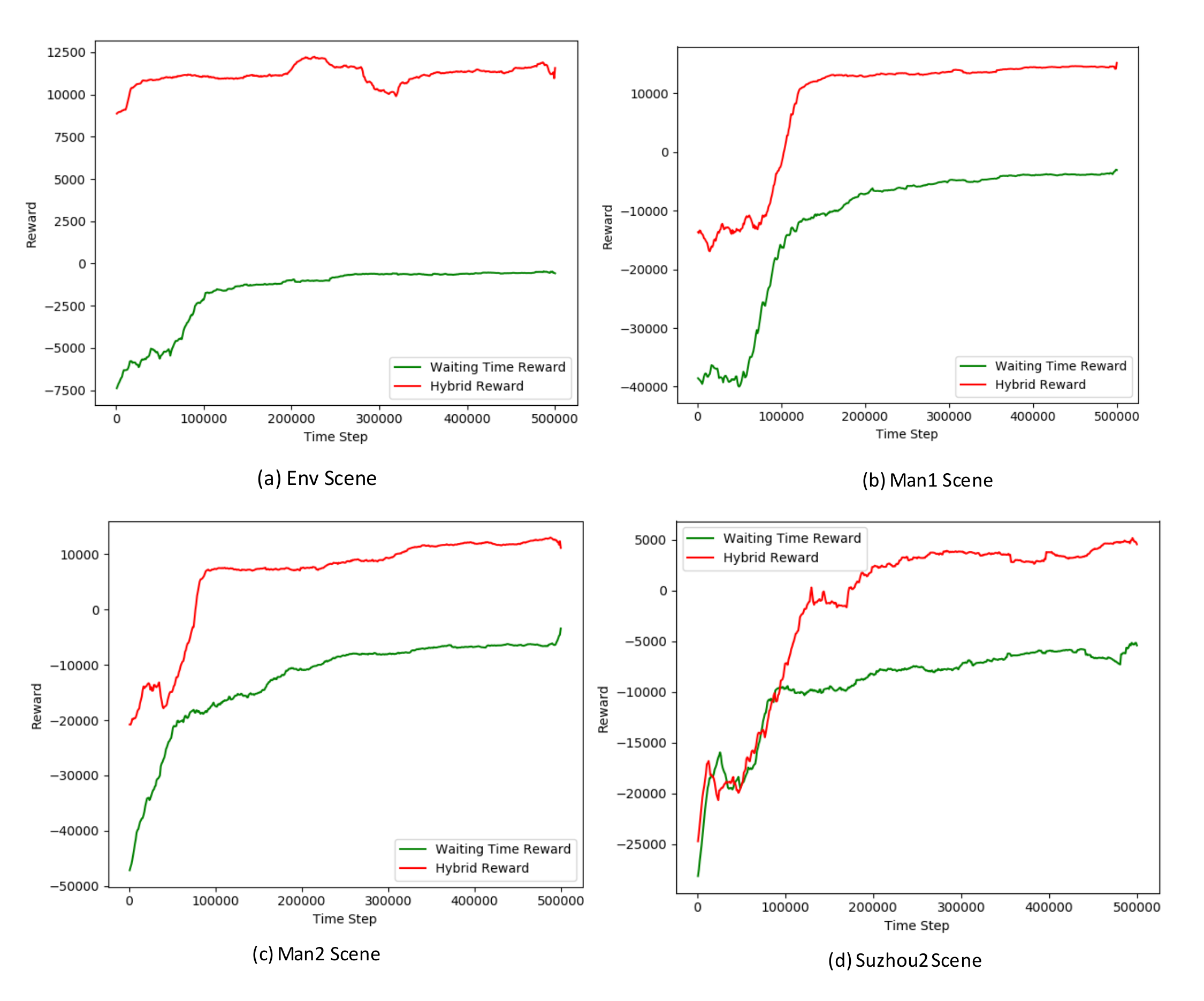}
\caption{Comparison of EGU-RL with hybrid reward and with only waiting time reward under different scenes. The proposed algorithm can obtain higher reward with hybrid reward, which suggests the effectiveness of hybrid reward to practically control the frequency of traffic signals.}
\label{fig:unchanged_r}
\end{figure}

We have also compared the reward obtained by EGU-RL with hybrid reward and with only total accumulated waiting time reward. Empirically, the reward trade-off factor $\alpha$ is set to 1.0. Figure \ref{fig:unchanged_r} suggests that with hybrid reward, the proposed model is able to receive much more reward. Consequently, it represents the effectiveness of hybrid reward to practically control the frequency of traffic signals.

\HU{We further conduct extra experiments on vehicles’ stop times. In SUMO, the vehicles’ stop time returns the consecutive time in where this vehicle was standing (voluntary stopping is excluded). Therefore, we sum up the stop waiting time of each vehicle as the evaluation. The results are listed in the table \ref{tab:stop-time}}

\begin{table}[t]
\centering
\caption{Cross-model vehicles’ stop time consumption comparison in 1000 steps. The unit here is hours. The less vehicles’ stop time consumed, the better performance of the model has. The EGU-RL method achieves the best results among all the other methods in most cases.}
\scalebox{1.2}{
\setlength{\tabcolsep}{0.85mm}{
\begin{tabular}{l|ccc|cc}
\hline
Scenes           & Fix  & Auction \cite{DBLP:journals/corr/BalujaCS17} & MARL-s       & EGU-RL       \\ \hline
Man2-flow1       & 1571  & 1049    & 712   & \textbf{390} \\ \hline
\end{tabular}}}
\label{tab:stop-time}
\end{table}

\HU{Similar to evaluation under accumulative waiting time of vehicles, the stop time optimised by the proposed EGU-RL is significantly lower than the competing methods, including Fix-schedule scheme, Auction-base method \cite{DBLP:journals/corr/BalujaCS17} and Single-agent for single-junction strategy. The EGU-RL model reduces more than 75\% of vehicle stop time compared with Fix-schedule scheme under the Man2-flow1 scenario with busy traffic flow. It also reduces ~50\% of vehicle stop time when compared with MARL-s.}

\subsection{EGU-RL with other Reinforcement Learning algorithms}

Our proposed EGU-RL model is flexible to combine with different reinforcement learning algorithms. Actor-Critic (AC) algorithm with EGE and USD further proves the point. The results are shown in Table \ref{tab:other_rl}.

\begin{table}[t]
\centering
\caption{The Actor-Critic (AC) algorithm with EGE and USD are shown as EGU-RL-AC. Performance comparison of EGU-RL-DQN and EGU-RL-AC on different environments.}
\setlength{\tabcolsep}{0.5mm}{
\begin{tabular}{l|cc}
\hline
Scenes              & EGU-RL-DQN           & EGU-RL-AC             \\ \hline
Env-flow1        & \textbf{222} & 269           \\
Env-flow2        & \textbf{86}  & 96            \\
Env-flow3        & 57           & \textbf{57}   \\
Man1-flow1  & 1978         & \textbf{1730} \\
Man1-flow2  & \textbf{549} & 555           \\
Man1-flow3  & 395          & \textbf{355}  \\
Man2-flow1  & 875          & \textbf{726}  \\
Man2-flow2  & \textbf{469} & 488           \\
Man2-flow3  & 321          & \textbf{289}  \\
Suzhou1-flow & 315          & \textbf{167}  \\
Suzhou2-flow1     & 843          & \textbf{793}  \\
Suzhou2-flow2     & \textbf{368} & 382           \\
Suzhou2-flow3     & \textbf{239} & 268           \\ \hline
\end{tabular}}
\label{tab:other_rl}
\end{table}

In Table \ref{tab:other_rl}, both EGU-RL-DQN and EGU-RL-AC models are able to achieve promising results on Env-flow1 environment.

\section{Conclusion}

In this paper, we have carefully designed  settings for machine learning based traffic control. We have also proposed new datasets, consists of one synthetic map and multiple real maps along with real traffic flows. We have proposed a novel and strong EGU-RL model to tackle multi-intersection traffic optimisation problems.

The proposed EGU-RL model is able to not only exploit multi-intersection spatial relations with an edge-weighted graph convolutional encoder, but also jointly models all controlled junctions with the unified structure decoder. The model integrates all junctions by outputting an action probability matrix with a unified structure decoder. By doing so, significantly fewer parameters are required to train and thus it reduces the overfitting risk. It facilitates networks to learn complex interactions of different intersections because the actions are jointly modelled. Experiments show that the performance achieved by the proposed model surpasses existing methods and the proposed model is more effective and practical in real traffic control scenes due to its simplicity.

\ifCLASSOPTIONcaptionsoff
  \newpage
\fi

\bibliographystyle{IEEEtran}
\bibliography{mybib.bib}

\end{document}